# METAL-GATED JUNCTIONLESS NANOWIRE TRANSISTORS

Mostafizur Rahman*, Pritish Narayanan and Csaba Andras Moritz

*Dept. of Electrical and Computer Engineering, University of Massachusetts Amherst, USA*
*rahman@ecs.umass.edu**

Junctionless Nanowire Field-Effect Transistors (JNFETs), where the channel region is uniformly doped without the need for source-channel and drain-channel junctions or lateral doping abruptness, are considered an attractive alternative to conventional CMOS FETs. Previous theoretical and experimental works [1][2] on JNFETs have considered polysilicon gates and silicon-dioxide dielectric. However, with further scaling, JNFETs will suffer from deleterious effects of doped polysilicon such as high resistance, additional capacitance due to gate-oxide interface depletion, and incompatibility with high-κ dielectrics[3][4]. In this paper, novel metal-gated high-κ JNFETs (MJNFETs – Fig. 1) are investigated through detailed process and device simulations. These MJNFETs are also ideally suited for new types of nano-architectures such as N$^3$ASICs [5] which utilize regular nanowire arrays with limited customization. In such nano-systems, the simplified device geometry in conjunction with a single-type FET circuit style [6] would imply that logic arrays could be patterned out of pre-doped SOI wafers without the need for any additional ion implantation.

In the MJNFET, the material properties of the gate and channel wires in conjunction with the nanoscale dimensions enable FET-like switching characteristics without the need for engineered source/drain junctions or lateral doping abruptness. Depending on the work function of gate, the heavily doped channel region of junctionless transistor is fully depleted without the application of an external bias (normally-off device – Fig. 2A). As the gate voltage increases, the electron concentration in the channel region starts to increase until it reaches the doping concentration $N_d$, at which point the flat-band condition is reached (Figs. 2B – 2D).

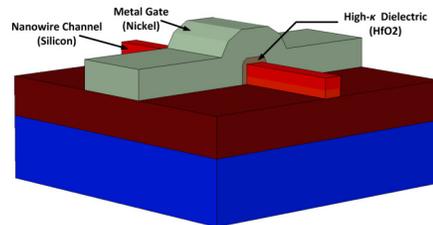

Fig. 1 Schematic diagram of junctionless transistor on SOI

The workfunction requirement for the gate in MJNFETs is tuned to achieve full-channel depletion. Materials with higher (lower) workfunction than the doped semiconductor channel are expected to deplete n-type (p-type) channels. Fig. 3 shows the range of metal workfunctions[7] suitable for achieving n- and p-type FETs. This is in direct contrast to workfunction requirements for conventional FETs [3] (e.g. n-type MOSFETs would require workfunction < 4.6).

The properties of MJNFETs were analyzed through detailed TCAD simulations [8]. The device structures were created using Synopsys Process considering detailed process effects such as implantation parameters, diffusion temperature, oxide growth etc. Given the nanowire dimensions under consideration, full 3-D simulations of the device structure are required. A quantum confinement model was used for modeling charge transport. The channel dimensions were 22nm (length) X 10nm (width) X 10nm (height). Channel doping was assumed to be n-type (2 X 10$^{19}$ dopants/cm$^3$) to achieve a sufficiently high on-current. HfO$_2$ gate dielectric with 2nm thickness was assumed. Metal workfunctions between 4.63 (Tungsten) to 5.22 (Nickel) was assumed. Higher workfunctions would imply very high threshold voltages. Lower workfunctions would imply poor off-state depletion for the nanowire dimensions and doping considered.

Fig. 4 shows simulated characteristics of MJNFETs. $I_d$ – $V_d$ curves (Fig. 4A) show saturation effects with increasing $V_d$. This is due to pinch-off of the channel near the drain region similar to inversion-mode devices. There is also a linear decrease in the on-current with increasing metal workfunction (Fig. 4B) with currents between 6µA and 21µA, comparable to on-currents of conventional inversion-mode devices. Fig. 4C shows a linear increase in threshold voltage between 0.2V to 0.8V with increasing workfunction ($V_{th}$ trend similar to inversion-mode devices discussed in [3]), as well as excellent on/off current ratios. Compared to a polysilicon gate with very high doping (Fig. 4D), higher on/off current ratios are achieved for the same threshold voltage (e.g. 100X better at $V_{th}$=0.4V), implying better subthreshold swing. These results imply that MJNFETs meet device requirements for integration both into conventional CMOS and the N$^3$ASIC nano-architecture. By suitably adjusting workfunction, doping levels and oxide parameters, characteristics can be tuned to meet specific technology requirements. More details on process and device simulation methodologies, capacitance extraction, results and expanded discussion will be included in the full paper.

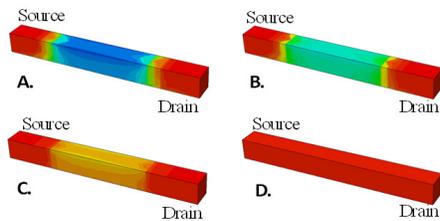

Fig. 2 **Carrier concentration contour plot in n-type junctionless transistor. A-D**, A.Full depletion at $V_G$=0, B. At $V_G$=$V_{TH}$ narrow channel formation, C. At $V_G$ > $V_{TH}$ carrier accumulation, D. At $V_G$ = $V_{FB}$ full carrier accumulation

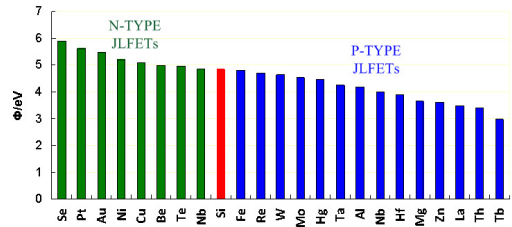

Fig. 3 Ideal metal work functions (clean surface)

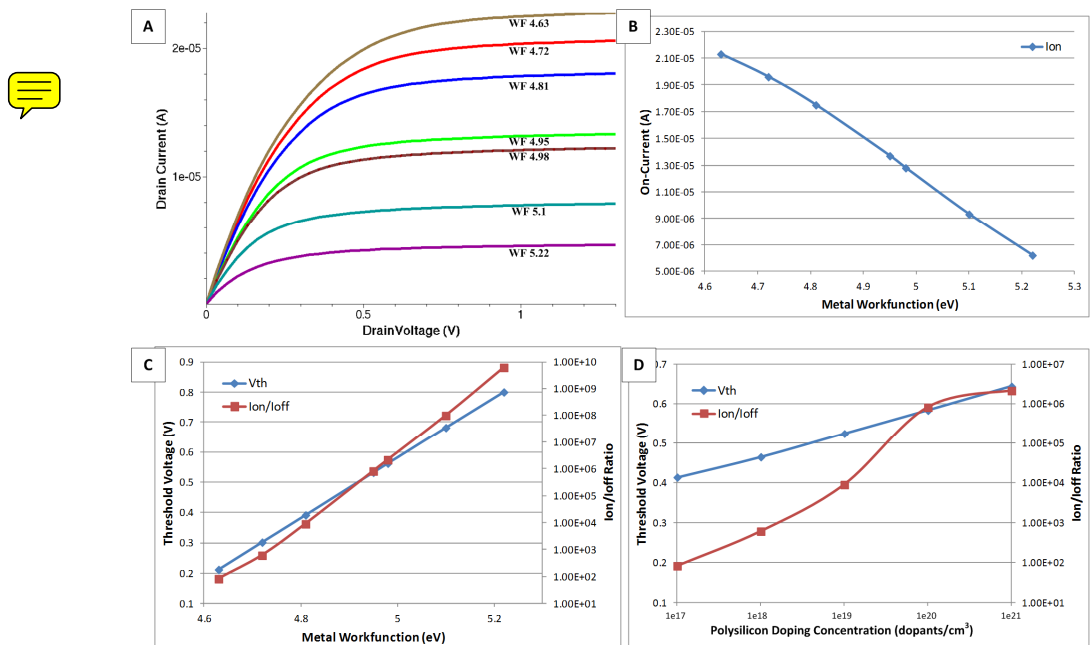

Fig. 4. A) Simulated $I_d$-$V_d$ characteristics of MJNFETs, , A. $I_d$-$V_d$ curve, B. $I_{ON}$ Voltage trend, C. Threshold Voltage and on/off ratios for MJNFETs D) Voltage and on/off ratios for polysilicon JNFETs.